# ULTRAFAST DYNAMICS OF GOLD NANORODS: TUNING BETWEEN PHOTO-BLEACHING AND PHOTO-INDUCED ABSORPTION

M. ANIJA,[1] SUNIL KUMAR,[1] N. KAMARAJU,[1] NEHA TIWARI,[2] S. K. KULKARNI,[2] A. K. SOOD[1,*]

[1]Department of Physics and Center for Ultrafast Laser Applications, Indian Institute of Science, Bangalore 560 012, India

[2]Indian Institute of Science Education and Research, First Floor, Central Tower, Sai Trinity Building, Garware Circle, Sutarwadi, Pashan, Pune 411 021, India

*asood@physics.iisc.ernet.in

We report ultrafast electron dynamics in gold nanorods investigated using 80 fs laser pulses centered at 1.57 eV. Five types of nanorod colloidal suspensions in water having their longitudinal surface plasmon peak ($E_{LSP}$) on either side of the laser photon energy ($E_L$) have been studied. For $E_{LSP} > E_L$, photo-induced absorption with single decay time constant is observed. On the other hand, for $E_{LSP} < E_L$, photo-bleaching is observed having bi-exponential decay dynamics; the faster one between 1-3 ps and slower one between 7ps to 22ps both of them increasing almost linearly with the difference $|E_L - E_{LSP}|$. These time constants increase linearly with the pump intensity. Simulations have been carried out to understand the interplay between photo-bleaching and photo-induced absorption.

## 1. Introduction

Metallic nanorods are of special interest due to tunability of their longitudinal surface plasmon (LSP) resonance from visible to near infrared. Gold nanorods have been used recently in multiplexed optical recording to increase information density beyond 1 TB cm$^{-3}$ [1] and for therapeutical applications [2]. Fundamental properties of nanorods heavily depend on their electron-electron and electron-phonon couplings which have been selectively addressed by using femtosecond time resolved differential transmission measurements [3,4]. Using pump at 400 nm and probe at 700 nm, photo-bleaching signal from gold nanorods having LSP resonance peak at 750 nm with electron-phonon relaxation time of ~ 3ps was observed [3]. Degenerate pump-probe experiments at 790 nm in resonance with the LSP peak also showed photo-bleaching with electron-phonon relaxation time varying from 1.5 ps to 10 ps as the pump intensity was varied from 0.5 GW/cm$^2$ to 12 GW/cm$^2$ [4].

Here, we have studied electron relaxation dynamics in different gold nanorod suspensions in water having aspect ratio from 3 to 5 and LSP resonance peak varying from 660 nm to 849 nm using 80 fs laser pulses in degenerate differential transmission measurements at 790 nm (1.57 eV). The change of sign of the differential transmission signal from photo-bleaching to photo-induced absorption for samples with LSP peak on either side of the laser photon energy is observed and explained. Simulations show that the pump pulse creates athermal excess electron distribution above Fermi level that changes the metal dielectric function followed by relaxation in two time scales. The aspect ratio of the nanorods decides the position of the LSP peak and photo-bleaching or photo-induced absorption is observed depending on whether LSP peak energy ($E_{LSP}$) is lesser or higher with respect to the probe photon energy ($E_L$).

## 2. Experimental Procedure

Gold nanorods with aspect ratio ranging from 3 to 5 are prepared by seed mediated growth technique as described by Nikoobakht *et al.* [5]. Seed solution is formed by mixing 0.2M CTAB solution with 0.5mM HAucl$_4$ followed by stirring for a few minutes. This is followed by adding 0.6ml of ice-cold 10mM NaBH$_4$ and vigorously stirring for two minutes [6]. Growth solution is prepared by adding 0.2M CTAB to 4mM AgNO$_3$ in deionized water. To this, 1mM HAucl$_4$ is added and after gentle mixing of the solution, 70ml of 78.8 mM ascorbic acid is added. Finally, 12 μl of the seed solution is added to the growth solution for the formation of gold nanorods. Nanorods with different aspect ratio were obtained by varying the amount of AgNO$_3$ in the Growth solution. Fig. 1 shows linear absorption spectra of four such samples. The transverse surface plasmon resonance peak occurs at ~ 525 nm (2.4 eV) but the longitudinal surface plasmon resonance peak varies from 660 nm to 849 nm ($E_{LSP}$ = 1.88 eV to 1.46 eV); accordingly the samples are named following their LSP peak position.

## 3. Results

The differential transmission signals ($\Delta T/T$) from four nanorod samples are shown in Fig. 2. The results have been simulated theoretically as described in the following section and are shown by black lines. It can be seen that photo-bleaching is observed for samples when $E_{LSP}<E_L$ and photo-induced absorption when $E_{LSP}>E_L$. We first discuss the results of Fig. 2 obtained at low pump fluence of ~ 0.13mJ/cm$^2$. For the photo-bleaching case, bi-exponential decay is observed with decay time constant $\tau_1$ and amplitude $A_1$ and decay time constant $\tau_2$ (amplitude 1-$A_2$): the time constants and amplitudes increase linearly with the energy difference $E_{LSP}$-$E_L$ as shown in Fig. 3. As the LSP band maximum moves away from the laser photon energy, the time constants increase significantly. On the other hand, photo-induced absorption signal decays mono-exponentially with a time constant $\tau_1$ of ~ 3ps, irrespective of the energy difference $E_{LSP}$-$E_L$.

The effect of pump fluence on the relaxation times is shown in Fig. 4 for two samples: AuNR849 ($E_{LSP}<E_L$) and AuNR760 ($E_{LSP}>E_L$).

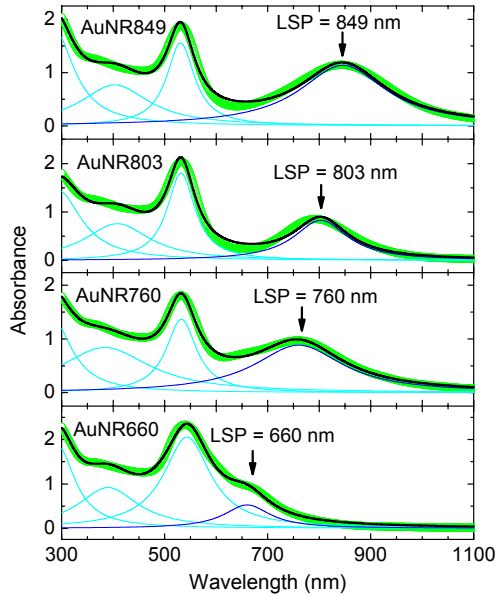

Fig. 1. Linear absorption spectra of gold nanorod colloidal suspensions in water for four samples having longitudinal surface plasmon resonance varying from 660 nm to 849 nm. Data are shown by open circles and thick black line is the fit using four Lorentzians (thin lines).

Degenerate pump-probe measurements were carried out using 80 fs laser pulses centered at 790 nm ($E_L$ = 1.57 eV) taken from an amplifier (1 kHz, Spitfire, Spectra Physics). The experiments were performed on gold nanorod colloidal suspensions in water having LSP peak centered at 660nm, 760nm, 802nm, 803nm, 806 nm and 849 nm. Pump fluence was varied from 0.13mJ/cm$^2$ to 1.0mJ/cm$^2$, and probe was kept constant at 8μJ/cm$^2$. The change in transmitted probe intensity in presence of the pump was recorded as a function of time delay between the two pulses in a usual lock-in detection scheme.

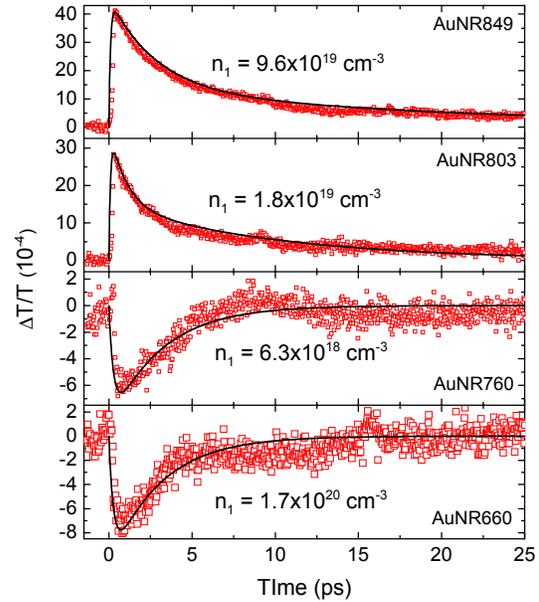

Fig. 2. Differential transmission signal from gold nanorod colloidal suspensions at pump fluence of 0.13mJ/cm$^2$. Solid lines are the fits obtained from the simulation as discussed in the paper. The excess carrier density ($n_1$) generated by the pump is given for each sample.

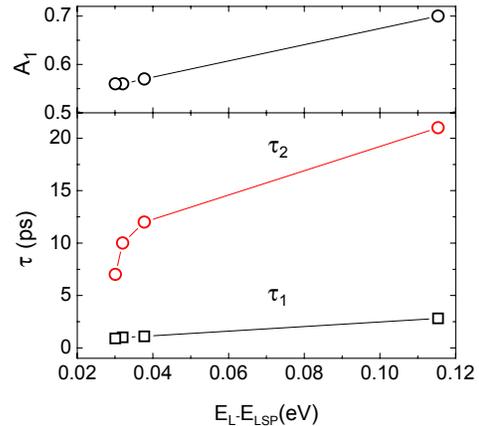

Fig. 3. Parameters of the bi-exponential dynamics of photo-bleaching signal as a function of position of the longitudinal surface plasmon peak ($E_{LSP}$) with respect to laser photon energy ($E_L$).

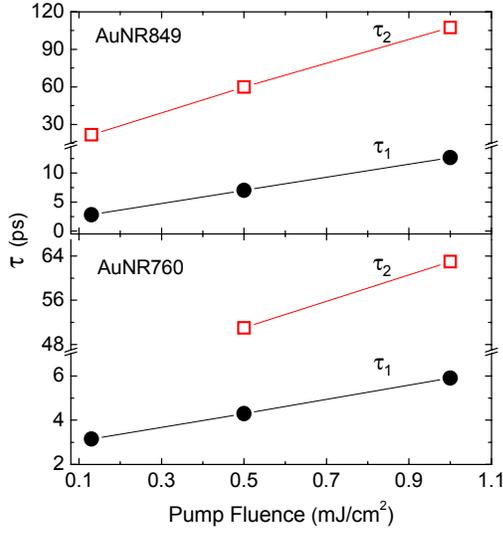

Fig. 4. Comparison of the carrier relaxation dynamics in AuNR849 and AuNR760 as a function of pump fluence.

## 4. Discussion

In gold the interband transition energy is ~ 2.4 eV [7], in close proximity with the transverse surface plasmon peak but quite off from the LSP peak (1.46 to 1.88 eV). Hence degenerate pump-probe measurements at 1.57eV can investigate only the electron distribution by intraband absorption. It has been shown in silver nanoclusters that when excited with a laser pulse either in resonance or out of resonance from the surface plasmon peak, the induced coherent electron density oscillations dephase within a time scale of ~ 10 fs [8,9] and do not significantly affect differential transmission of the probe. Only quasi-free electron excitation is observed to be a dominant contributor [10]. Hence in describing the conduction band electron dynamics in metal films, nanoparticles and nanorods, only the incoherent effects need to be taken into account. Thermal redistribution of energy among the photo-excited electrons takes place through electron-electron scattering and a new hot Fermi distribution is set-up. After this thermalization, energy is transferred to the lattice through electron-phonon coupling and to the surrounding matrix through phonon-phonon interaction at a much slower relaxation rate.

To understand the photo-bleaching and photo-induced absorption depending on the LSP peak energy with respect to the laser photon energy, we have carried out a detailed analysis of electronic photo-excitation in gold nanorods. The electronic response of nanoparticles with size > 10 nm can be considered to follow the bulk dielectric function [11,12]. Energy injection to the conduction band electrons by the pump pulse modifies the Drude contribution to the metal dielectric function. As a result, the absorption at probe energy ($E_L$) is modified [13,14]. The change in probe transmission can be calculated as a function of time delay between pump and probe pulses and hence the electron dynamics is captured.

The linear absorption coefficient of gold nanorod colloidal suspension can be written as [15,16,17]

$$\alpha(\omega) = p \frac{\varepsilon_d^{3/2} \omega}{3c} \sum_{j=1}^{3} \frac{(1/P_j^2)\varepsilon_2(\omega)}{\varepsilon_2^2(\omega) + [\varepsilon_1(\omega) + (1/P_j - 1)\varepsilon_d]^2} \quad (1)$$

Here $p$ is the volume fraction of nanorods, $\varepsilon_d = 1.77$ is the dielectric constant of surrounding water [18], $c$ is speed of light in vacuum, and $P_j$ ($j = 1,2,3$) are the depolarizing factors along three axes of nanorods (prolate ellipsoids). If $P_1$ is considered to be along the long axis then

$$P_1 = (1/e^2 - 1)\left[(1/2e)\ln\left(\frac{1+e}{1-e}\right) - 1\right]; \; e = \sqrt{1 - 1/\xi^2} \quad (2)$$

and $P_2 = P_3 = (1-P_1)/2$. $\xi$ is the aspect ratio of the nanorods and decides the position of LSP peak position in the absorption spectrum as can be seen from Eq. 1. $\varepsilon_1$ and $\varepsilon_2$ are the real and imaginary parts of the complex dielectric function of gold and can be written as sum of contributions from the free electrons (Drude part) and bound electrons (interband) given by

$$\varepsilon_1(\omega) + i\varepsilon_2(\omega) = 1 - \frac{\omega_p^2}{\omega^2 + i\Gamma\omega} + \frac{\widetilde{\omega}_p^2}{\omega_0^2 - \omega^2 - i\gamma\omega} \quad (3)$$

where $\omega_p = (ne^2/\varepsilon_0 m_e)^{1/2} \sim 13.8 \times 10^{15}$ sec$^{-1}$ is the plasma frequency, $\Gamma \sim 1.05 \times 10^{14}$ sec$^{-1}$ is the electron scattering rate [19]. The instantaneous electron density, $n$ depends on the pump intensity. For bound electrons, $\widetilde{\omega}_p \sim 4.5 \times 10^{15}$ sec$^{-1}$ is analogous to plasma frequency, $\gamma \sim 9 \times 10^{14}$ sec$^{-1}$ and interband transition energy $\omega_0 = 2.4$ eV [19]. Now the differential transmission signal of the probe is simply followed from Beer Lambert's law and can be written as,

$$\frac{\Delta T}{T} = -\Delta(\alpha L) \quad (4)$$

$L$ is the sample optical path length. Eq. (4) can be rewritten at probe photon energy ($E_L = \hbar\omega_{pr}$) as

$$\left.\frac{\Delta T}{T}\right|_{\omega=\omega_{pr}} = -\Delta\alpha(\omega_{pr})L = -L\frac{\partial\alpha}{\partial\varepsilon_1}\bigg|_{\omega=\omega_{pr}}\Delta\varepsilon_1 - L\frac{\partial\alpha}{\partial\varepsilon_2}\bigg|_{\omega=\omega_{pr}}\Delta\varepsilon_2 \quad (5)$$

The pump laser pulse creates intensity dependent non-equilibrium carrier density, $n_1$ which relaxes back to zero and can be monitored at each time delay $t$ between the pump and the probe pulses. The instantaneous electron density can be modeled as:

$$n(t) = n_0 + n_1[1-\exp(-t/\tau_R)]\sum_i A_i \exp(-t/\tau_i) \quad (6)$$

Here $n_0$ is the equilibrium electron density, $\tau_R$ is the rise time taken by athermal electrons to thermalize after pump excitation, $A_i$ are amplitudes and $\tau_i$ are various relaxation times. The time dependent plasma frequency $\omega_p = [n(t)e^2/\varepsilon_0 m_e]^{1/2}$ is substituted into Eq. (3) and hence the absorption coefficient can be calculated from Eq. 1. By varying $p$ and $\xi$, LSP frequency is matched with the measured values for different samples. Finally the differential transmission signal is calculated from Eq. (5) at each time delay $t$ using $n_1$, $\tau_R$, $A_i$ and $\tau_i$ as the fitting parameters. Good fits were obtained using only two decay components ($A_1$ and $A_2 = 1-A_1$, $\tau_1$ and $\tau_2$). $n_1$ for different samples are given along with the fits shown by black lines in Fig. 2. For photo-bleaching, $\tau_R \sim 100$ fs and for photo-induced absorption $\sim 300$ fs was obtained at all pump fluences. It can be seen that the model can explain quantitatively photo-bleaching for $E_{LSP}<E_L$ and photo-induced absorption for $E_{LSP}>E_L$.

## 5. Conclusion

In conclusion, we have studied electron dynamics in gold nanorods having their longitudinal surface plasmon resonance varying on either side of the laser photon energy. Photo-bleaching or photo-induced absorption is observed depending on whether $E_{LSP}<E_L$ or $E_{LSP}>E_L$. The photo-generated excess carrier density depends on the pump intensity and relaxes more slowly at higher intensities. Simulations have been carried out to understand photo-bleaching vis-a-vis photo-induced absorption based on a dielectric model.

### Acknowledgments

A.K.S. thanks Department of Science and Technology, India for financial support and S.K. acknowledges University Grants Commission for Senior Research Fellowship.